\documentclass{aa}
\usepackage{txfonts}
\usepackage[utf8]{inputenc}
\usepackage{graphicx}
\usepackage{amssymb}
\usepackage{amsmath}
\usepackage{subfig}
\usepackage[colorlinks=true,citecolor=blue,linkcolor=blue]{hyperref}

\newcommand{\ppf}{\texttt{PreProFit\;}}
\newcommand{\ppfnospace}{\texttt{PreProFit}}

\begin{document}

\title{PreProFit -- Pressure Profile Fitter for galaxy clusters}
\author{Fabio Castagna \and Stefano Andreon}
\institute{INAF–Osservatorio Astronomico di Brera, via Brera 28, 20121 Milano, Italy, \email{fabio.castagna@inaf.it}}

\abstract{
Galaxy cluster analyses based on high-resolution observations of the Sunyaev--Zeldovich (SZ) effect have become common in the last decade.
We present \ppfnospace, the first publicly available code designed to fit the pressure profile of galaxy clusters from SZ data. \ppf is based on a Bayesian forward-modelling approach, allows the analysis of data coming from different sources, adopts a flexible parametrization for the pressure profile, and fits the model to the data accounting for Abel integral, beam smearing, and transfer function filtering.
\ppf is computationally efficient, is extensively documented, has been released as an open source Python project, and was developed to be part of a joint analysis of X-ray and SZ data on galaxy clusters. 
\ppf returns $\chi^2$, model parameters and uncertainties, marginal and joint probability contours, diagnostic plots, and surface brightness radial profiles. 
\ppf also allows the use of analytic approximations for the beam and transfer functions useful for feasibility studies.}
\maketitle

\section{Introduction}
Galaxy clusters are the largest and most massive gravitationally bound objects in the Universe, and thus they offer a unique tracer of cosmic evolution.
The thermodynamic properties of a galaxy cluster can be gathered from observation in the optical band, the X-ray band, or microwaves.
The hot gas trapped in the cluster's gravitational potential leaves an imprint on the microwave sky because its electrons Compton scatter the photons of the cosmic microwave background radiation \citep{Sunyaev1970, Sunyaev1972}. The distortion is best observable at millimetre wavelengths and is directly proportional to the pressure distribution in the clusters.
Specifically, the amplitude of the SZ effect is parametrized as the Compton $y$ parameter.

The number of high-resolution SZ instruments has progressively increased throughout the last decade, including the
NIKA\footnote{New IRAM KIDs Array} 
camera \citep{Monfardini2010} and the 
GISMO\footnote{Goddard IRAM Superconducting Millimiter Observatory} 
camera \citep{Staguhn2008} at the 
IRAM\footnote{Institut de Radio Astronomie Millimétrique} 
30m telescope, the 
MUSTANG\footnote{Multiplex SQUID TES Array at Ninety GHZ} 
camera \citep{Dicker2008} on the 100m Robert C. Byrd Green Bank Telescope, the \textit{Planck} satellite \citep{Planck2013}, the Bolocam array \citep{Czakon2015} and the 
MUSIC\footnote{Multiwavelength Submillimeter kinetic Inductance Camera} 
camera \citep{Sayers2010} on the Caltech Sub-millimeter Observatory, and the 
ALMA\footnote{Atacama Large Millimeter/submillimeter Array}+ACA\footnote{Atacama Compact Array} 
and 
CARMA\footnote{Combined Array for Research in Millimeter-wave Astronomy} 
\citep{Woody2004} arrays.
Furthermore, a new generation of instruments such as NIKA2 \citep{Calvo2016}, MUSTANG2 \citep{Dicker2014}, and TolTEC \citep{Austermann2018} have recently been developed, which have improved the quality and quantity of SZ data.

Several studies on the SZ effect have been performed \citep[e.g.][]{Birkinshaw2005, Mroczkowski2009, Korngut2011, Sayers2013, Adam2015, Romero2017} and the methodologies for operating with these data are constantly evolving.
In most cases, the SZ data analysis improves in order to meet the specific demands of each analysis (e.g. using the actual beam in place of a Gaussian approximation of it).

In this paper, we present \ppfnospace, which is, to the best of our knowledge, the first publicly available code for fitting the pressure profile of galaxy clusters.
\ppf is meant to automate and generalize all the phases of data analyses in an efficient and easy-to-use software pipeline, and therein lies its most remarkable feature.

\ppf includes highly time-consuming operations such as convolutions. As a result, our purpose throughout the software development was to find a way to perform these tasks sufficiently quickly, but without losing accuracy. 
\ppf is able to analyse data coming from different sources and also allows the use of analytic approximations for the beam and transfer functions, useful for feasibility studies, or to make use of published data with only approximate information on theses quantities.
\ppf has been developed to be part of a joint analysis of X-ray and SZ data on galaxy clusters.

The paper is organised as follows: in Sect.~\ref{sec:ppf} we provide an overview of the software, as well as the technical requirements; in Sect.~\ref{sec:methods} we describe in detail the methodology behind each step of the program; and in Sect.~\ref{sec:example} we present an application of \ppf on real data from the galaxy cluster CL~J1226.9+3332. We conclude with the discussion and final remarks in Sect.~\ref{sec:conclusion}.

\begin{figure*}[tbp]
    \centering
    \includegraphics[height=10cm]{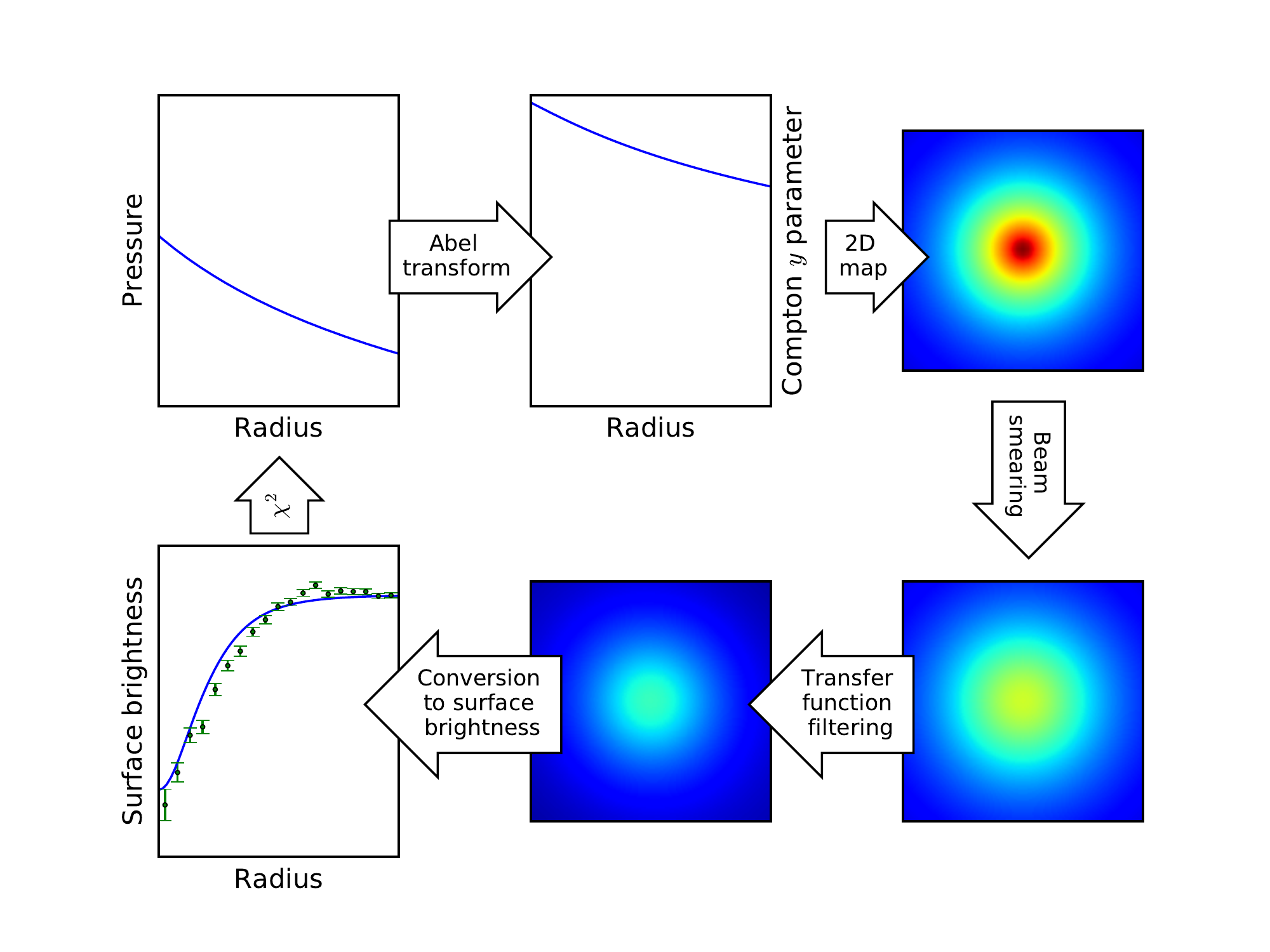}
    \caption{Block diagram showing the program flow.}
    \label{fig:pipeline}
\end{figure*}

\section{\ppfnospace} \label{sec:ppf}

\subsection{Program flow}
\ppf adopts a flexible parametrization for the pressure profile of the cluster and properly derives the surface brightness profile to be compared with the observed data. It assumes spherical symmetry for the cluster, as in most analyses \citep[e.g.][]{Comis2011, Adam2015, Romero2015, Romero2017}. As outlined in Fig.~\ref{fig:pipeline}, \ppf  projects the three-dimensional pressure profile into a two-dimensional map using the forward Abel transform. The map is then convolved with the instrumental beam and the transfer function.
With opportune conversion factors, we finally derive the surface brightness profile, whose fit to the data is measured through the likelihood function of the model. \ppf automatically produces a number of diagnostic plots, including the model and data radial surface brightness profile. 

\subsection{Requirements and installation}
\ppf was developed and tested with Python 3.6. The following libraries were required to build: mbproj2, PyAbel, numpy, scipy, astropy, emcee, six, matplotlib, corner. 
\ppf can be downloaded from GitHub\footnote{https://github.com/fcastagna/preprofit}.

\section{Methods} \label{sec:methods}
\subsection{Key stages}
To follow, we present a step by step description of \ppf working principles, as illustrated in Fig.~\ref{fig:pipeline}.
\subsubsection{Pressure profile}
The pressure profile is described by the generalized Navarro, Frenk \& White (gNFW) model proposed by \cite{Nagai2007}:
\begin{equation}
    P_e(r)=\frac{P_0}{\left(\frac{r}{r_p}\right)^c\left(1+\left(\frac{r}{r_p}\right)^a\right)^{\frac{b-c}{a}}},
    \label{eq:press_prof}
\end{equation}
where $P_0$ is a normalizing constant and $r_p$ is a scale radius. The exponentials $b$ and $c$ describe the logarithmic slopes at $r/r_p\gg1$ and $r/r_p\ll1$, respectively, while $a$ governs the rate of turnover between these two slopes. The five parameters make the model very flexible; they can fit current data and allow users to identify which parameters are constrained by the data.

\subsubsection{Abel integral}
The three-dimensional pressure model is numerically integrated along the line of sight in order to obtain a two-dimensional map of the Compton $y$ parameter. This is performed according to the Abel transform:
\begin{equation}
    y(R)=\frac{\sigma_T}{m_e c^2}\int_{R}^{R_b}\frac{2r P_e(r)}{\sqrt{r^2-R^2}}dr,
\end{equation}
where $\sigma_T$ is the electron Thompson scattering cross section, $m_e$ is the electron rest mass, $c$ is the speed of light, and $R_b$ is the cluster radial extent. 
To calculate the integral, \ppf makes use of the well-developed Python function \textit{direct\_transform} from the PyAbel library.

\begin{figure*}[htbp]
    \centering
    \subfloat{{\includegraphics[height=6cm]{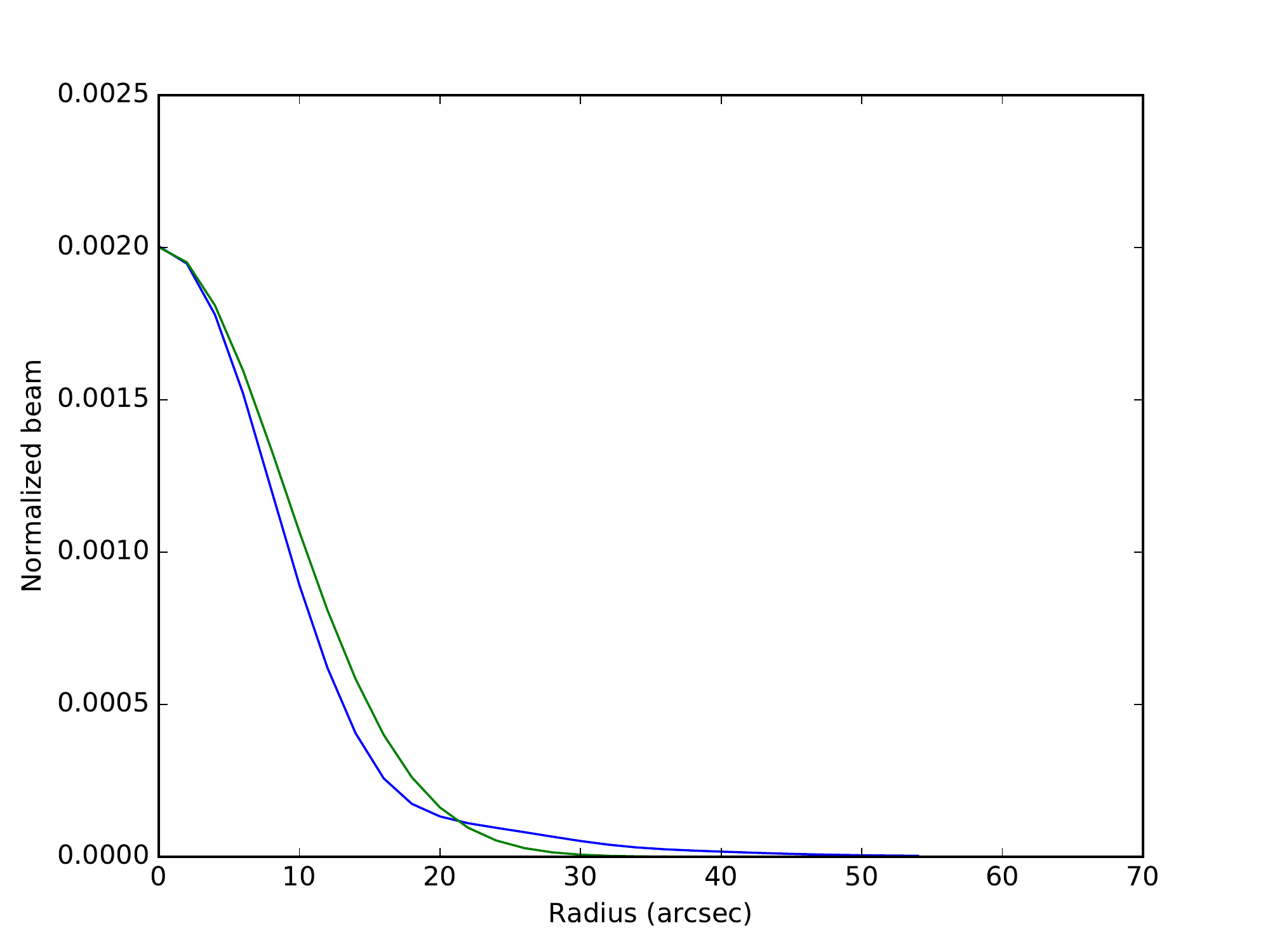}}}
    \qquad
    \subfloat{{\includegraphics[height=6cm]{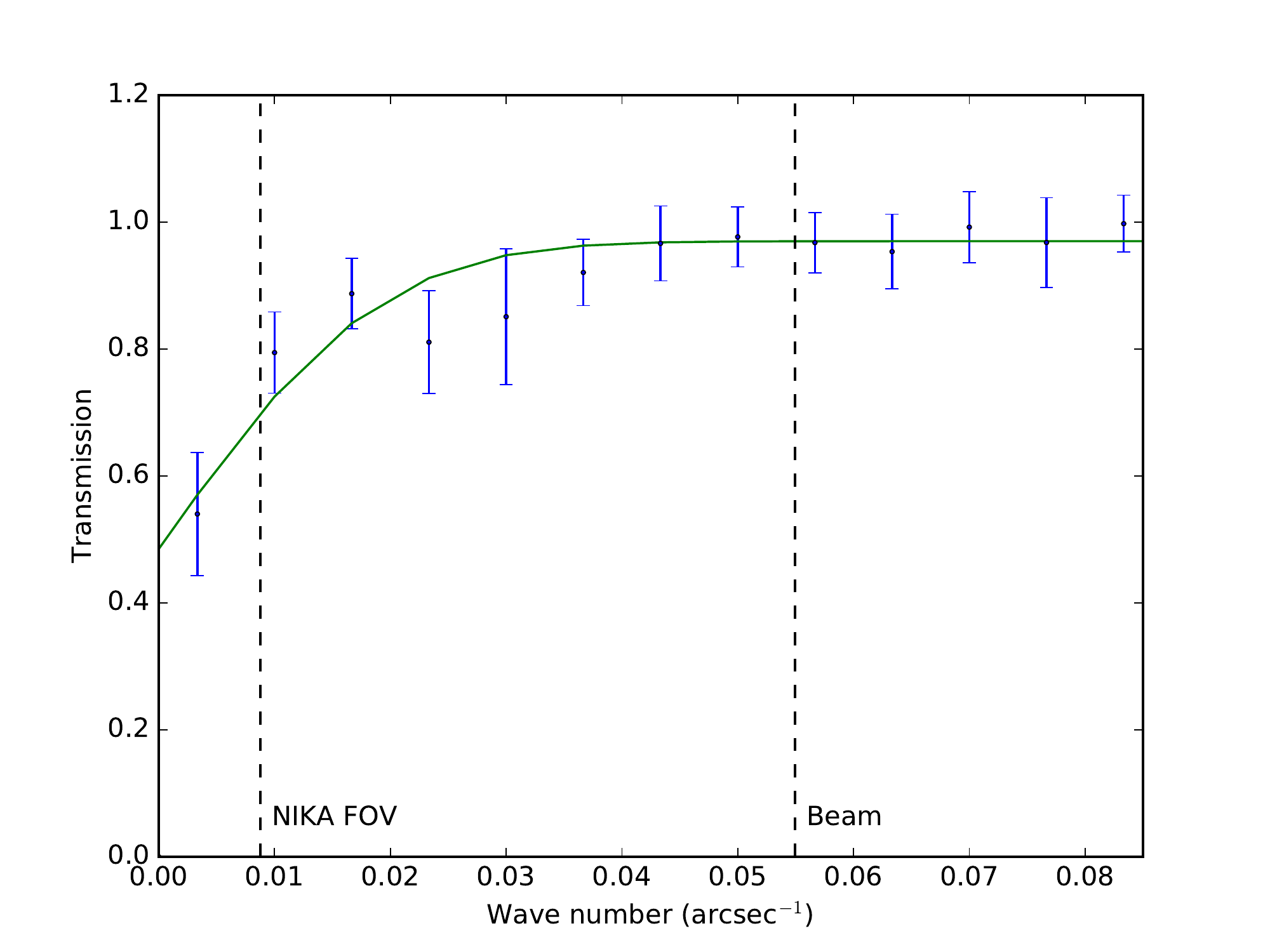}}}
    \caption{\textit{Left panel}: Beam profile. \textit{Right panel}: Transfer function profile. Measured profiles are in blue, the approximations used for Fig.~\ref{fig:approx}  in green.}
    \label{fig:nika_prof}
\end{figure*}

At this stage, we create a two-dimensional map for the Compton $y$ parameter profile adopting a regular grid and complying with radial symmetry.

\subsubsection{Beam smearing} \label{subsec:beam}
In order to consider the effects of the instrument used for observing the cluster, we need to convolve our model map with the beam.
\ppf supports either to read the beam data from a file or to approximate it with a Gaussian distribution (e.g. based on the instrumental full width at half maximum) when, for example, these data are not available. By default, the beam map is normalized to have an integral of one, but it can even be left as such on user request. The convolution is performed via fast Fourier transform (FFT).

\subsubsection{Transfer function}
After convolving with the beam, the input transfer function\footnote{The transfer function (transmission) is often obtained as the square root of the ratio of the one-dimensional power spectra of the observed fake sky and input fake sky \citep{Romero2017}.} is applied to the model in the Fourier space.
If requested, \ppf allows the use of an approximation based on the cumulative distribution function of a Gaussian with location, scale, and normalization parameters chosen by the user.
From the radial transfer function profile, we construct a two-dimensional transmission image assuming radial symmetry.
To pass our model through the signal transfer function, we multiply the unfiltered map in the Fourier space with the 
transmission image just created. 
Compared to implementations in  \cite{Adam2015} and \cite{Romero2017}, \ppf adopts a more general approach for the transfer function filtering allowing free values for the image pixel size and the transfer function sampling step.

\subsubsection{Conversion to surface brightness}
To compare our model map with the observed data, the filtered Compton $y$ parameter map is converted into surface brightness, measured in Jy/beam, using the input conversion factor.

\subsection{Model definition}
The whole set of operations described in the previous sections represents our model, which allows us to derive a surface brightness map from a parametrized pressure profile. From the surface brightness map, we extract the radial profile and compare it with the measured data, thus setting up the likelihood function of the model,  the probability of observing the data given the model parameters. As in previous
analyses \citep[e.g.][]{Adam2015, Ruppin2017}, we adopt the likelihood function:
\begin{equation}
\mathcal{L}=\exp\left(-\frac{\chi^2}{2}\right),
\end{equation}
where
\begin{equation}
    \chi^2=\sum_{i=1}^n\left(\frac{f_i^{data}-f_i^{model}}{\sigma_i^{data}}\right)^2.
    \label{eq:likelihood}
\end{equation}
Here $f^{data}$ and $f^{model}$ are the observed and the estimated surface brightness value, respectively, while $\sigma^{data}$ is the error on the data and $n$ is the number of available data points.

\subsection{Markov chain Monte Carlo algorithm}
The posterior is sampled with an affine-invariant ensemble sampler proposed by \citet{Goodman2010} and implemented in
\textit{emcee} \citep{Foreman2013}.
The user has to specify the list of parameters to be fitted, and optionally can change the bounds of each parameter's prior uniform distribution.
The user is free to fix the desired number of random walkers, the number of iterations, and the burn-in period extent, as well as the starting values of the chains.
Multi-threading computation is supported by \ppf and is strongly encouraged to minimize the time of execution.

Qualitative and quantitative diagnostics are both provided by \ppf to evaluate whether the chains adequately converge  to the stationary distribution. First of all, the acceptance fraction is automatically displayed in the program output. Next, \ppf allows  the traceplot and the cornerplot to be reproduced, which inform about the evolution of the parameter values across the iterations and the joint posterior distribution, respectively. Finally, the user can visualize the best (median) fitting surface brightness profile, together with its confidence interval (CI), and compare it to the observed data. The $\chi^2$ value and the number of degrees of freedom are also given.

\begin{figure*}[htbp]
    \centering
    \subfloat{{\includegraphics[height=7cm]{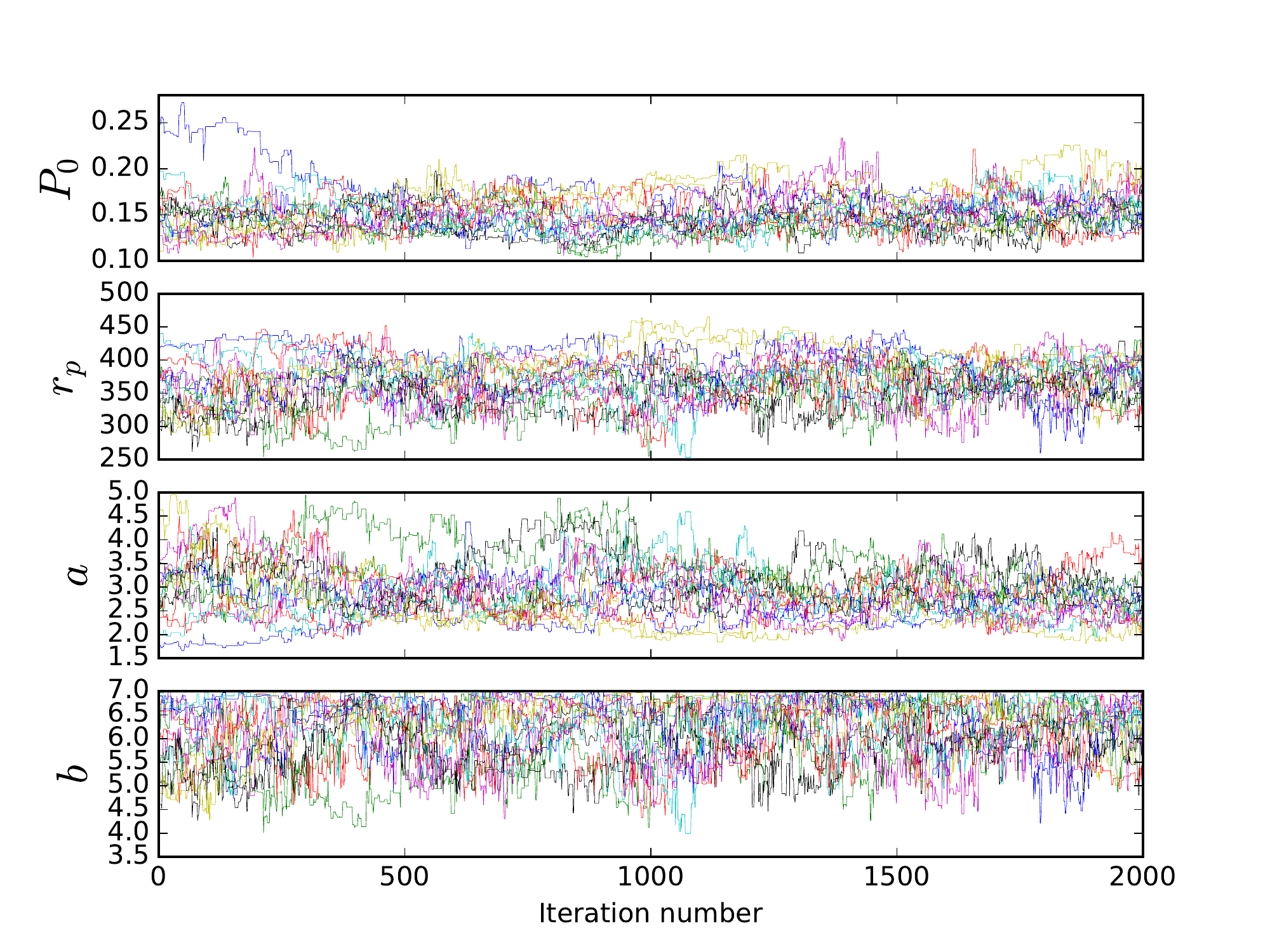}}}
    \qquad
    \subfloat{{\includegraphics[height=7cm]{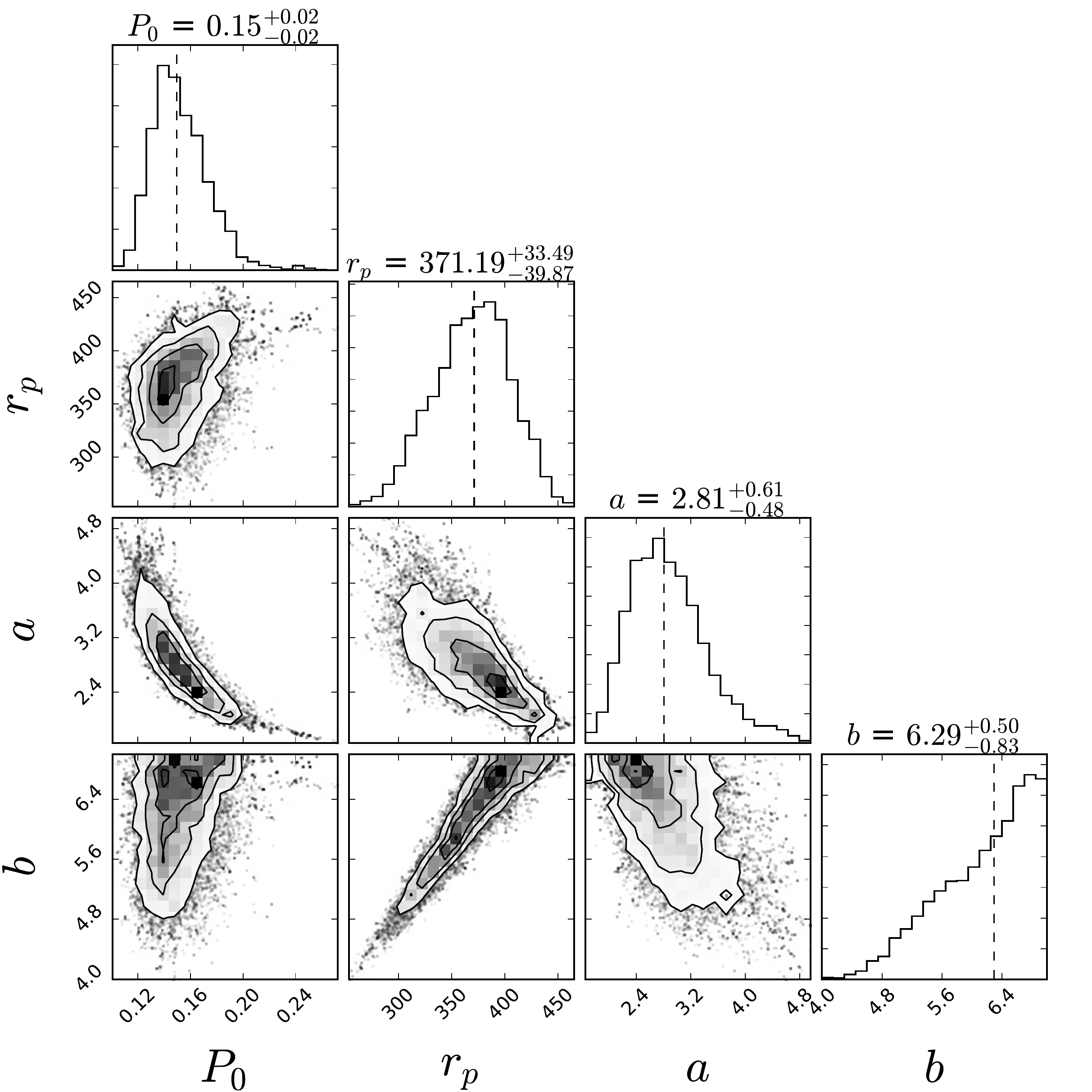}}}
    \caption{Diagnostic plots automatically produced by \ppfnospace. \textit{Left panel}: Trace plot. \textit{Right panel}: Joint and marginal posterior distributions.}
    \label{fig:diagnostics}
\end{figure*}

\subsection{Validation}
The whole processing pipeline has been validated in its entirety.
The \textit{direct\_transform} function, which calculates the Abel integral, is approximately 50 times faster than the common \textit{quad} function from the Scipy library \citep{SciPy} and keeps its accuracy within 0.8\% of the true values.
The \textit{fftconvolve, fft2, ifft2} functions, used for convolution, and direct and inverse Fourier transform computation, respectively, come from the well-known Scipy Python library. To the best of our knowledge, they are the most computationally efficient functions for performing such operations. The
\textit{fftconvolve} function is about 100 times faster than the standard \textit{convolve2d} function, also from the Scipy library, and reports a $<10^{-7}$\% approximation error. 
As an additional test of the beam smearing (see Sect.~\ref{subsec:beam}), we compared the performance of \textit{fftconvolve} with a direct convolution using MIDAS \citep{Banse1983}: our implementation is more than ten times faster, with a systematic error lower than $10^{-7}$\%. We even checked the computational accuracy in the case of a convolution between two Gaussians whose solution is analytic: the relative difference is below $10^{-7}$\%\ again.

\subsection{Execution time}

The fit of pressure profile can be extremely slow \citep[e.g.][]{Ruppin2019} because of the Abel integral, beam convolution, and transfer function filtering.
By working on the different steps, we achieved a balanced load: Abel transform requires 33\% of the CPU time, two-dimensional image interpolation 18\%, beam smearing 24\%, transfer function filtering 23\%, and other minor operations account for the remaining 2\%.

\begin{figure}[!pb]
    \centering
    \includegraphics[height=7cm]{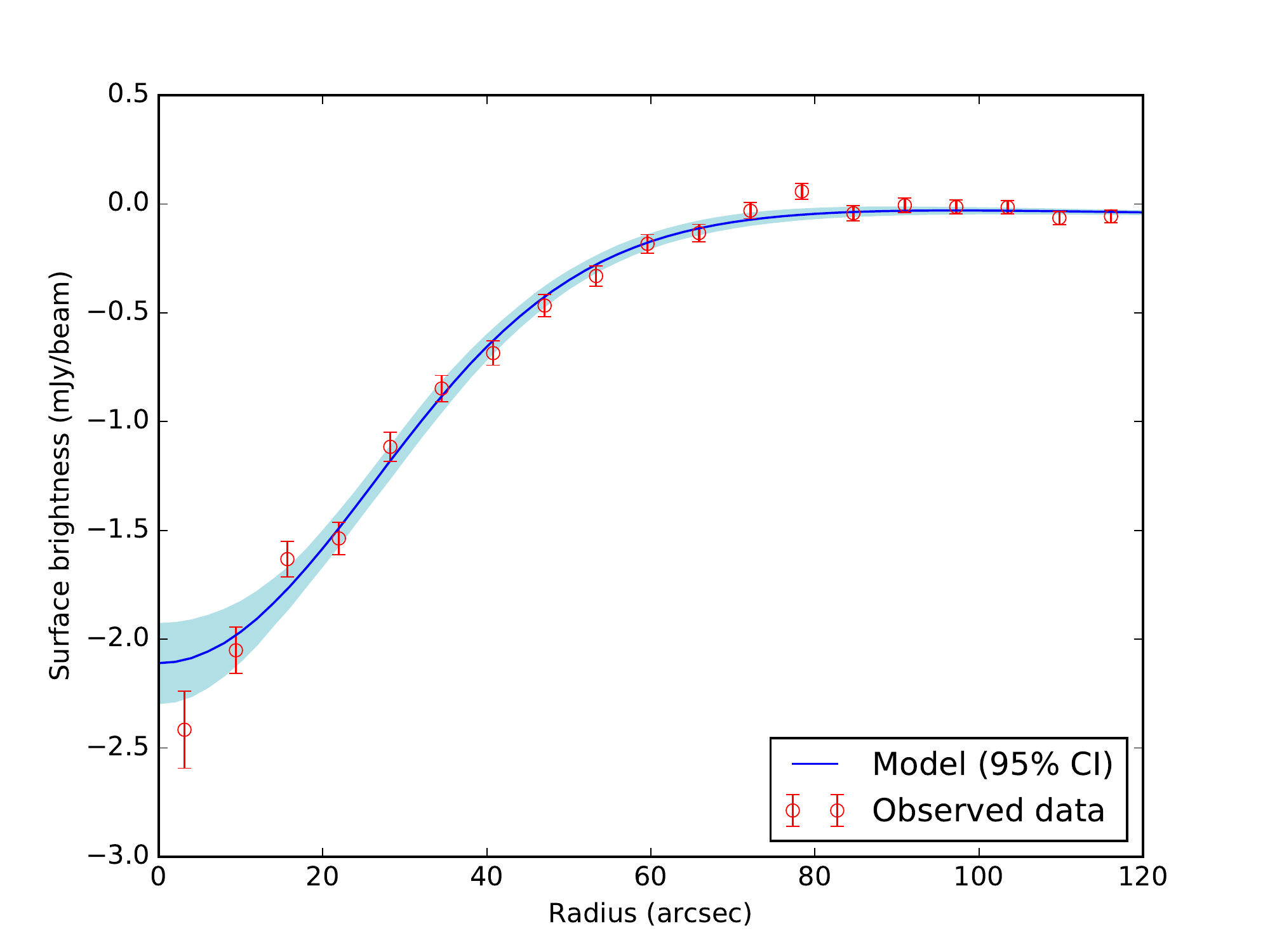}
    \caption{Surface brightness profile (points with 68\% error bars) and best fitting profile with 95\% CI. This plot is automatically generated by \ppfnospace.}
    \label{fig:bestfitprof}
\end{figure}

\begin{figure}[!ht]
    \centering
    \includegraphics[height=7cm]{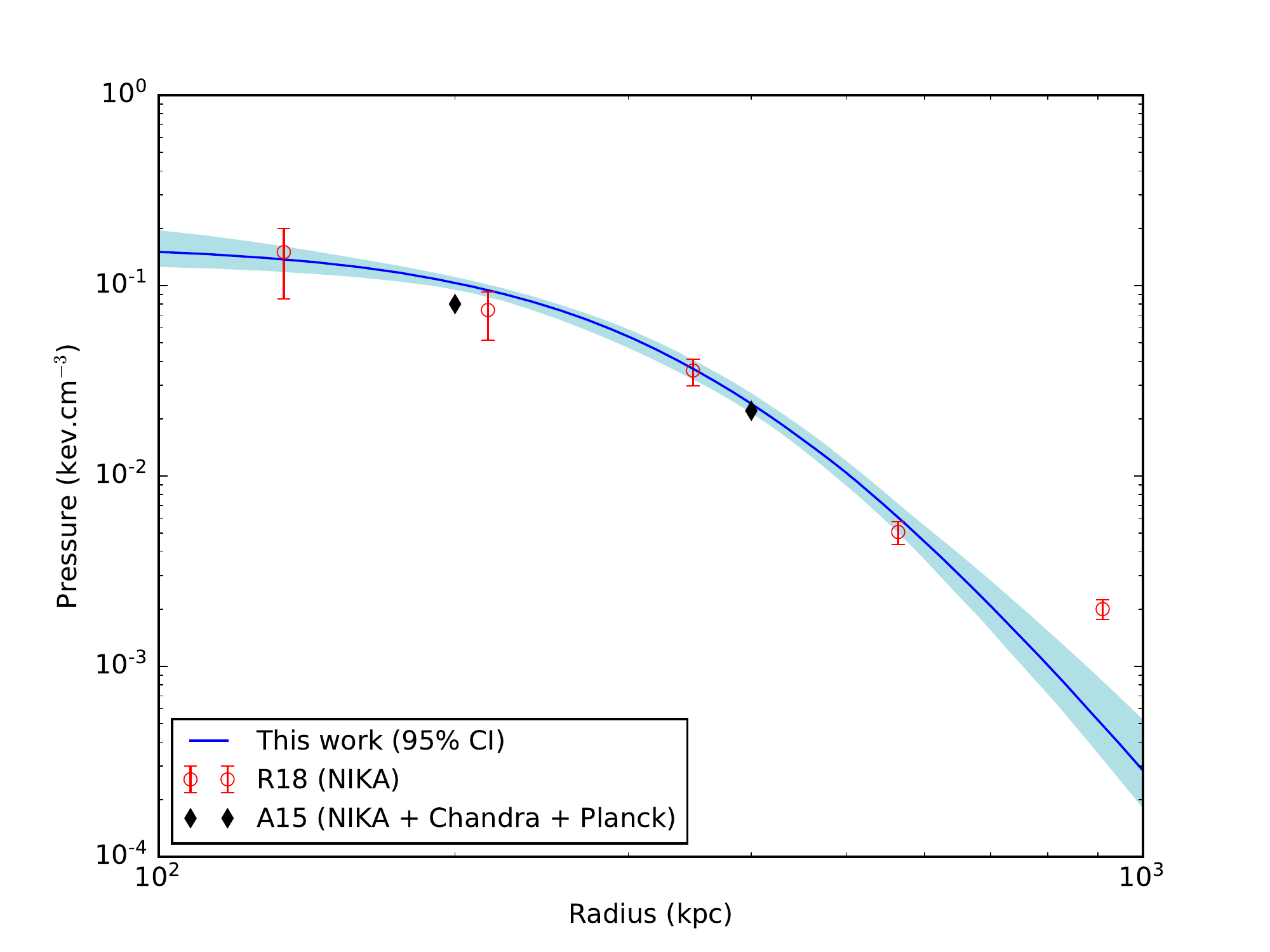}
    \caption{Comparison of CL~J1226.9+3332 pressure profiles. Our best fit is plotted in blue (95\% CI shaded); the red points are from the non-parametric fit of \cite{Romero2018} on NIKA data (68\% error bars); the black diamonds without error bars refer to the fit of \cite{Adam2015} on NIKA, \textit{Chandra}, and \textit{Planck} data.}
    \label{fig:adamplot}
\end{figure}

\section{A worked example} \label{sec:example}
To highlight the potentiality of \ppfnospace, we present an application of the program on real data. We chose to analyse the high-redshift cluster of galaxies CL~J1226.9+3332 ($z=0.89$), which has been largely studied by several authors in recent years, both in the X-ray \citep{Maughan2004, Maughan2007, Donahue2014} and in SZ \citep{ Mroczkowski2009, Korngut2011, Sayers2013, Adam2015, Romero2017, Romero2018}. CL~J1226.9+3332 is a hot and massive cluster, discovered in 2001 with the \textit{ROSAT} WARPS survey \citep{Ebeling2001}. The cluster presents a relaxed morphology on a large scale, with some possible evidence of a disturbed core \citep{Maughan2007, Korngut2011, Adam2015, Romero2017}. In our work we flagged the point source detected by \cite{Adam2015} at (RA, Dec) = (12:26:59.855,+33:32:35.21).

\subsection{NIKA}
The SZ observation, instrumental beam, transfer function, and spectral conversion coefficient used in the example come from the publicly available NIKA data release\footnote{http://lpsc.in2p3.fr/NIKA2LPSZ/nika2sz.release.php}. We refer the reader to \cite{Adam2015} for more details about the cluster observation, even though we remark how the data on the site were reduced and filtered in a slightly different way. In Fig.~\ref{fig:nika_prof} we show the beam profile and the transfer function, as well as their normal and cumulative normal approximations,   to illustrate the behaviour of \ppf when using such approximations.

\subsection{Model definition}
By leaving all the gNFW parameters in Equation~\ref{eq:press_prof} free, we found that  parameter $c$ is unconstrained by the data; therefore, we fixed $c=0.014$ as in \cite{Adam2015} and let the other  parameters free to vary.
As priors we took
\begin{multline*}
    \Theta=\big\{\left(P_0, r_p, a, b\right)\in \mathbb{R}^{4}: 0<P_0<1,\, \\
    100<r_p<1000,\, 0.5<a<5,\, 3<b<7 \big\},
\end{multline*}
where $P_0$ is expressed in keV cm$^{-3}$ and $r_p$ in kpc.

We ran 5000 iterations of 14 chains and considered the first 3000 as the burn-in period.
We extracted the starting values of the chains from a Gaussian distribution with $\mu=\left(0.4, 300, 1.33, 4.13\right)$ and $\sigma=0.1$ for all parameters.
We considered a sampling step of 2 arcsec and a cluster radial extent of 5 Mpc for the Abel integral computation. The Compton $y$ to Jy/beam conversion factor is taken from \citet{Adam2015}. We adopt a flat $\Lambda$CDM cosmology with ${H_0=67.32}$~km~s$^{-1}$~Mpc$^{-1}$, ${\Omega_M=0.3158}$, and ${\Omega_\Lambda=0.6842}$ \citep{Planck2018}.

\subsection{Results}
Trace plot and joint plus marginal posterior distributions, which are automatically produced by \ppfnospace, are shown in Fig.~\ref{fig:diagnostics}. The trace plot suggests that the chains adequately explored the parameter space and this regular behaviour is confirmed by the acceptance fraction value equal to 0.43. 
As expected, $b$ is largely degenerate with $r_p$.
Figure~\ref{fig:bestfitprof}, which is automatically generated as well, shows the best-fit profile and its 95\% CI superimposed on the observed data. We found ${\chi^2=24.2}$ for 15 degrees of freedom.

\begin{figure}[!pb]
    \centering
    \includegraphics[width=\linewidth]{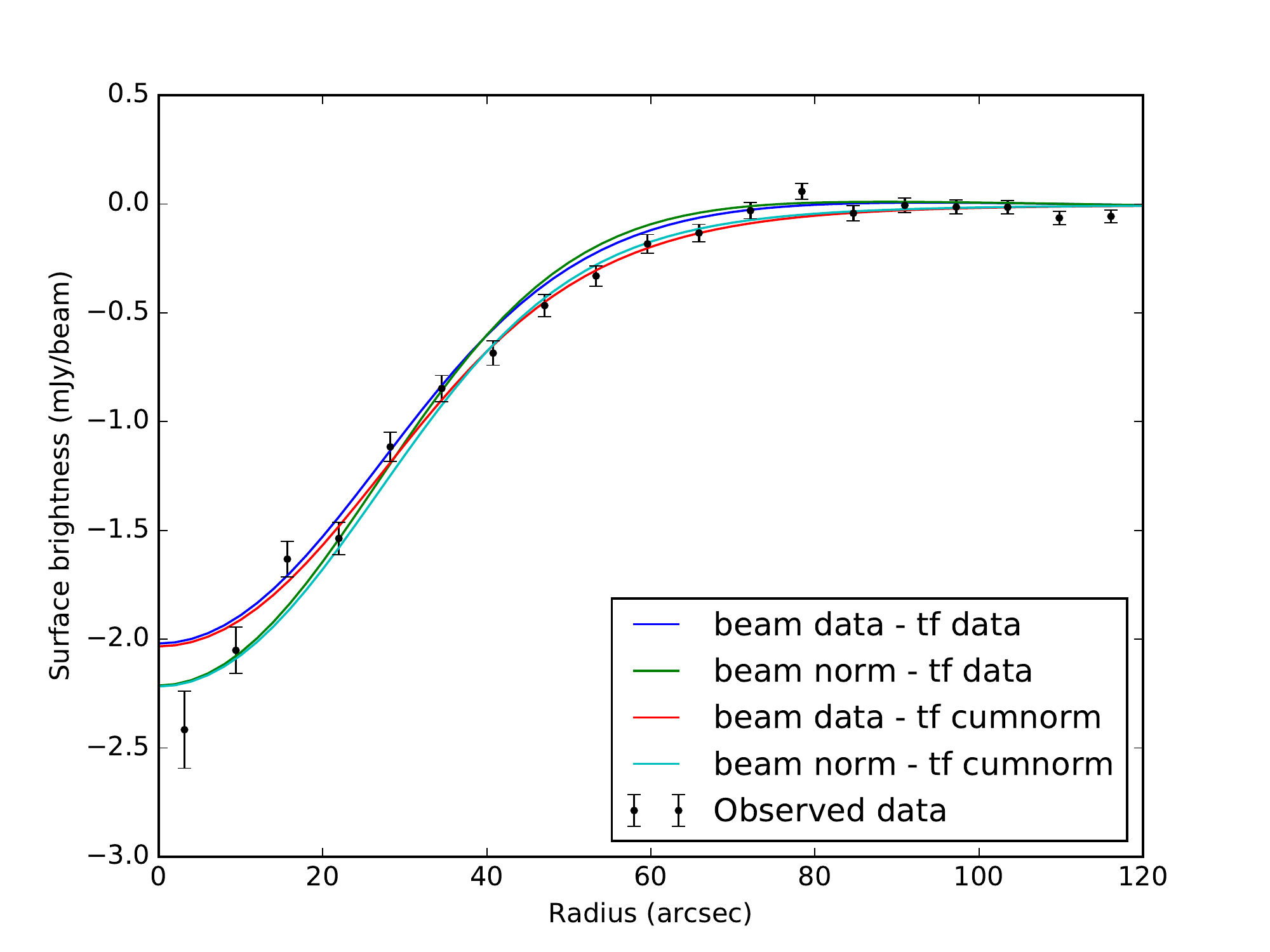}
    \caption{Surface brightness profiles derived for various combinations of beam and transfer function, as specified in the inset, for a single set of pressure profile parameters. Points with 68\% error bars show the data.}
    \label{fig:approx}
\end{figure}

Figure~\ref{fig:adamplot} compares our pressure profile against corresponding published results from \cite{Romero2018} and \cite{Adam2015} using the same NIKA data, even though they are differently reduced, and therefore have a different transfer function. 
The former made a non-parametric fit, which requires an unknown and uncertain deprojection correction of the outermost point, not included in the error budget. The latter added \textit{Planck} and \textit{Chandra} data and performed a joint fit: we read values without errors from their fig.~8. There is a good agreement among the three derivations, except at scales corresponding to the instrument radial field of view (around 1 Mpc), where derivations are uncertain, as is clearly documented in the literature \citep[e.g.][]{Sayers2016, Romero2019}.

In Fig.~\ref{fig:approx}, we demonstrate how close or different the results from either measured or approximated instrumental data turn out to be. Differences across surface brightness profiles are comparable to observed errors, indicating that approximated beam and transfer functions might be suitable for feasibility studies, but are unlikely to be useful for science analyses for data of this quality.

\section{Conclusion} \label{sec:conclusion}
The amount of data collected through SZ observations has regularly increased in recent years.
We have introduced a Python program, \ppfnospace, that allows users to estimate the pressure profile of galaxy clusters through flexible and efficient modelization.
\ppf is the first publicly available code to perform this kind of analysis and, notably, it is extensively documented;  by allowing the analysis of data coming from different sources, it can be useful to a wide community.
\ppf relies on a Bayesian forward-modelling approach. 

Users are free to set up \ppf in accordance with their needs and requirements. Among other things, users can decide how many and which parameters to fit, which (uniform) prior to adopt for the free parameters, and whether to conduct a feasibility study using approximations for the beam and the transfer function. \ppf returns $\chi^2$, model parameters and uncertainties, marginal and joint probability contours, diagnostic plots, and surface brightness radial profiles. 

After describing in detail each stage of the data processing pipeline, we  presented an application of the program on the high-redshift galaxy cluster CL~J1226.9+3332, using SZ observations from the NIKA camera. We outlined the main plots and Bayesian diagnostics produced by \ppf. We noted that, as is documented in the literature \citep[e.g.][]{Sayers2016, Romero2019}, it is very difficult to constrain the pressure profile near or beyond the instrument radial field of view.

The release of \ppf lays the groundwork for an enhanced version of the code that allows us to join the SZ analysis to a three-dimensional  (RA, Dec, energy) analysis of X-ray data to perform a powerful joint SZ-plus-X analysis.

\begin{acknowledgements}
We thank the referee for the useful comments on the manuscript.
F.C. acknowledges financial contribution from the agreement ASI-INAF n.2017-14-H.0 and PRIN MIUR 2015 Cosmology and Fundamental Physics: Illuminating the Dark Universe with Euclid.
\end{acknowledgements}

\bibliographystyle{aa}
\bibliography{references}

\end{document}